%% file: main_all.tex
\RequirePackage[mathlines]{lineno} 
\documentclass[a4paper,11pt]{article}
\pdfoutput=1 

\usepackage{jheppub} 

\usepackage[T1]{fontenc} 

\usepackage{threeparttable}
\usepackage{multirow}
\usepackage{subfigure}
\usepackage{float}
\let\oldequation\equation
\let\oldendequation\endequation
\renewenvironment{equation}
  {\linenomathNonumbers\oldequation}
  {\oldendequation\endlinenomath}

\def \jpsi {J/\psi}
\def \dmunv  {D^-\mu^+\nu_{\mu}}
\def \ee   {e^+e^-}

\def \gev  {~\mbox{GeV}}
\def \gevc {~\mbox{GeV/$c$}}
\def \gevcc{~\mbox{GeV/$c^2$}}
\def \mev  {~\mbox{MeV}}

\begin{document}

\title{\boldmath Search for the semi-muonic charmonium decay $J/\psi\to D^{-}\mu^{+}\nu_{\mu}+c.c.$}

\collaboration{The BESIII collaboration}
\emailAdd{besiii-publications@ihep.ac.cn}

\abstract{
Using $(10087\pm44)\times10^{6}$ $J/\psi$ events collected with the BESIII detector at the BEPCII $e^+e^-$ storage ring at the center-of-mass energy of $\sqrt{s}=3.097~\rm{GeV}$, we present a search for the rare semi-muonic charmonium decay $J/\psi\to D^{-}\mu^{+}\nu_{\mu}+c.c.$. Since no significant signal is observed, we set an upper limit of the branching fraction to be $\mathcal{B}(J/\psi\to D^{-}\mu^{+}\nu_{\mu}+c.c.)<5.6\times10^{-7}$ at $90\%$ confidence level. This is the first search for the weak decay of charmonium with a muon in the final state.
}


\arxivnumber{2307.02165}

\maketitle
\flushbottom

\section{INTRODUCTION}
\label{sec:introduction}
\hspace{1.5em}

\begin{figure}[htbp] \centering
	\setlength{\abovecaptionskip}{-1pt}
	\setlength{\belowcaptionskip}{10pt}
	\includegraphics[width=8.0cm]{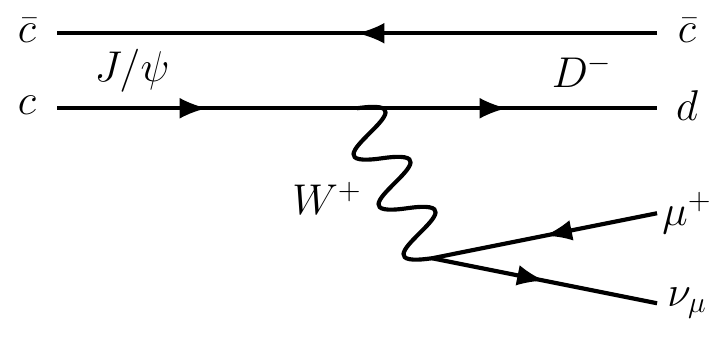}
	\caption{Feynman diagram for $\jpsi\to D^{-}\mu^{+}\nu_{\mu}$ decays at tree-level.}
	\label{fig:feynman}
\end{figure}

Since the discovery of the charmonium $\jpsi$ state in 1974~\cite{jpsi1:1974,jpsi2:1974}, its properties have been intensively studied. The $\jpsi$ decays primarily through strong and electromagnetic interactions. 
Because the mass of $\jpsi$ lies below the $D\bar{D}$ threshold, it cannot decay into a pair of charmed mesons,
but it is kinematically allowed for $\jpsi$ to decay weakly into a single charmed meson, for example $\jpsi\to D_{(s)}^{(*)} X$, where $X$ denotes a light hadron such as $\pi$ or a light lepton pair such as $e^{+}\nu_{e}$ or $\mu^{+}\nu_{\mu}$. 
Throughout this paper, charge-conjugate processes are implied.
The Feynman diagram of $\jpsi\to D^{-}\mu^{+}\nu_{\mu}$ at tree-level within the Standard Model (SM) is shown in Fig.~\ref{fig:feynman}.
\vspace{-0.0cm}
\vspace{-0.0cm}
In the SM, these rare weak decays of the $\jpsi$ into a single charmed meson are predicted to have an inclusive branching fraction (BF) of about $10^{-8}$ or below~\cite{verma:1990, Sanchis:1994, Sanchis:1993, sharma:1999, wang:2008a, wang:2008b, shen:2008, dhir:2013, ivanov:2015, tian:2017}, 
and none have yet been observed ~\cite{pdg:2022, bes:2008, bes3:2014xbo, bes3:2014, bes3:2017, bes3:2021}. The experimental results for the semi-leptonic channels are listed in Table~\ref{tab:exp}. 
However, there are some new physics theories, such as the Top-color Model~\cite{hill:1995}, the Minimal Supersymmetric SM with or without R-parity~\cite{Aulakh:1982yn}, and the two-Higgs doublet model~\cite{Glashow:1976nt}, in which these BFs could be significantly enhanced, reaching the level of $10^{-5}$~\cite{datta:1999}. 
Therefore, searches for charmonium weak decays reaching toward the SM predictions also probe new physics beyond the SM.

\begin{table*}[!htbp]
\caption{Experimental results for $\jpsi$  semi-leptonic weak decays. For each result, the experiment, the decay mode of $\jpsi$, the total size of the $\jpsi$ sample, and the upper limit (UL) on the BF at 90\% confidence level (C.L.) are given, along with the year and reference.}
\setlength{\abovecaptionskip}{1.2cm}
\setlength{\belowcaptionskip}{0.2cm}
\begin{center} 
\footnotesize
\vspace{-0.0cm}
\begin{tabular}{l|llll}
\hline \hline
		Experiment & Decay mode & $N_{\jpsi}$ & UL at 90\% C.L. & Year\\
		\hline
		BESIII & $J/\psi\to D_s^-e^+\nu_e$ & $225.3\times10^{6}$ & $1.3\times10^{-6}$ & 2014 \cite{bes3:2014}\\
		BESIII & $J/\psi\to D_s^{*-}e^+\nu_e$ & $225.3\times10^{6}$ & $1.8\times10^{-6}$ & 2014 \cite{bes3:2014}\\
		BESIII & $J/\psi\to D^0e^+e^-$ & $1310.6\times10^{6}$ & $8.5\times10^{-8}$ & 2017 \cite{bes3:2017}\\
		BESIII & $J/\psi\to D^-e^+\nu_e$ & $10087\times10^{6}$ & $7.1\times10^{-8}$ & 2021 \cite{bes3:2021}\\
\hline \hline
\end{tabular}
\label{tab:exp}
\end{center}
\end{table*}
\vspace{-0.0cm}

Thus far, the search for weak semi-leptonic charmonium decays has only covered the electron channel.
In the SM, charged leptons have identical electroweak interaction strengths, referred to as lepton flavor universality (LFU). 
But for $B^0\to D^{(*)-}l^+\nu_l$ ($l=e$, $\mu$, $\tau$), combined experimental results of $\mathcal{R}(D)$ and $\mathcal{R}(D^*)$ shows a tension with LFU at a significance of 3.2 standard deviations~\cite{Bigi:2016mdz,Bordone:2019vic,Gambino:2019sif,BaBar:2012obs,BaBar:2013mob,Belle:2015qfa,Belle:2019rba,Belle:2016dyj,LHCb:2015gmp,LHCb:2017smo,LHCb:2017rln,RD:sum}.
A search for the weak decay of charmonium with a muon in the final state is therefore desirable.
For the semi-muonic decay $\jpsi\to D^{-}\mu^{+}\nu_{\mu}$, 
the theoretical predictions within the SM are at the order of $10^{-11}$~\cite{wang:2008b, shen:2008, dhir:2013, ivanov:2015, tian:2017}, as shown in Table~\ref{tab:prediction}.

\begin{table*}[!htbp]
\caption{Theoretical results for the BF of the semi-leptonic decay $\jpsi\to D^{-}\mu^{+}\nu_{\mu}$ within the SM, where QCDSR is the QCD sum rule model, CLFQ is the covariant light-front quark model, BSW is the Bauer-Stech-Wirbel model, CCQM is the confined covariant quark model, and BSM is the Bathe-Salpeter-Mandelstam model.}
\setlength{\abovecaptionskip}{1.2cm}
\setlength{\belowcaptionskip}{0.2cm}
\begin{center}
\footnotesize
\vspace{-0.0cm}
\begin{tabular}{l|cccccccccc}
\hline \hline
                    Model & QCDSR~\cite{wang:2008b} & LFQM~\cite{shen:2008} & BSW~\cite{dhir:2013} & CCQM~\cite{ivanov:2015} & BSM~\cite{tian:2017}\\
                    \hline
                    BF $(\times10^{-11})$ 
                    & $0.71_{-0.22}^{+0.42}$ 
                    & $4.7-5.5$ 
                    & $5.8_{-0.6}^{+0.8}$ 
                    & $1.66$ 
                    & $1.98_{-0.24}^{+0.28}$\\
\hline \hline
\end{tabular}
\label{tab:prediction}
\end{center}
\end{table*}
\vspace{-0.0cm}

The BESIII~\cite{Ablikim:2009aa} experiment has collected $(10087\pm44)\times10^{6}$ $\jpsi$
events~\cite{bes3:njpsi2022} at a center-of-mass energy of $\sqrt{s}=3.097~\rm{GeV}$ operating at the Beijing Electron Positron Collider (BEPCII)~\cite{Yu:IPAC2016-TUYA01}. The experiment provides large samples to search for rare $\jpsi$ decays~\cite{li:2012, Ablikim:2019hff, Wang:2020gzc, Chen:2021fcb, BESIII:2022mxl}.
In this paper, we present the first search for the weak decay $\jpsi\to D^{-}\mu^{+}\nu_{\mu}$ with $D^{-}\to K^{+}\pi^{-}\pi^{-}$.
Pions can be misidentified as muons at low momentum~\cite{PID}, causing backgrounds not present for semi-electronic final states.  Since there are many $\jpsi$ decay modes with pions~\cite{pdg:2022}, the muon mode is more technically challenging. 
About $10\%$ of the full data sample is first used to validate the analysis procedure, and the final result is obtained from the full data sample with the validated analysis strategy.

\section{BESIII DETECTOR AND MONTE CARLO SIMULATION}
\label{sec:detector}
\hspace{1.5em}
The BESIII detector~\cite{Ablikim:2009aa} records symmetric $e^+e^-$ collisions 
provided by the BEPCII storage ring~\cite{Yu:IPAC2016-TUYA01} in the center-of-mass energy range from 2.0 to 4.95~GeV, with a peak luminosity of $1\times10^{33}\;\text{cm}^{-2}\text{s}^{-1}$ achieved at $\sqrt{s} = 3.77\;\text{GeV}$.
BESIII has collected large data samples in this energy region~\cite{Ablikim:2019hff}. The cylindrical core of the BESIII detector covers 93\% of the full solid angle and consists of a helium-based
 multilayer drift chamber~(MDC), a plastic scintillator time-of-flight
system~(TOF), and a CsI(Tl) electromagnetic calorimeter~(EMC),
which are 
all enclosed in a superconducting solenoidal magnet providing a 1.0~T magnetic field.
The magnetic field was 0.9~T in 2012, which affects 11\% of the total $J/\psi$ data.
The solenoid is supported by an
octagonal flux-return yoke with resistive plate counter muon
identification modules (MUC) interleaved with steel. 

The charged-particle momentum resolution at $1~{\rm GeV}/c$ is
$0.5\%$, and the ${\rm d}E/{\rm d}x$ resolution is $6\%$ for electrons
from Bhabha scattering. The EMC measures photon energies with a
resolution of $2.5\%$ ($5\%$) at $1$~GeV in the barrel (end cap)
region. The time resolution in the TOF barrel region is 68~ps, while
that in the end cap region is 110~ps. 
The end cap TOF system was upgraded in 2015 using multi-gap resistive plate chamber
technology, providing a time resolution of 60~ps, which benefits 87\% of the data used in this analysis~\cite{etof1, etof2}.
The minimum momentum of muon required for the MUC  to be effective is approximately 0.5 GeV/c. 

Simulated data samples produced with the {\sc geant4}-based~\cite{geant4} Monte Carlo (MC) package BOOST~\cite{bes:boost}, which includes the geometric and material description of the BESIII detector~\cite{detvis,geo1,geo2} and the detector response, are used to determine detection efficiencies and to estimate backgrounds. The simulation models the beam energy spread and initial state radiation (ISR) in the $e^+e^-$ annihilations with the generator {\sc kkmc}~\cite{ref:kkmc1, ref:kkmc2}. 
The inclusive MC sample includes the production of the $J/\psi$ resonance incorporated in {\sc kkmc}.
All particle decays are modeled with {\sc evtgen}~\cite{ref:evtgen1, ref:evtgen2} using the BFs either taken from the Particle Data Group~\cite{pdg:2022}, when available, or otherwise estimated with {\sc lundcharm}~\cite{ref:lundcharm1, ref:lundcharm2}.
Signal MC events are generated according to a $c\to d$ charged current weak interaction, similar to Refs.~\cite{bes3:2014, bes3:2021}.
Final state radiation~(FSR) from charged final state particles is incorporated using the {\sc photos} package~\cite{photos}.

\section{EVENT SELECTION AND DATA ANALYSIS}
\label{sec:analysis}
\begin{figure*}[htbp] \centering
	\setlength{\abovecaptionskip}{-1pt}
	\setlength{\belowcaptionskip}{10pt}
 
        \subfigure[]
        {\includegraphics[width=0.49\textwidth]{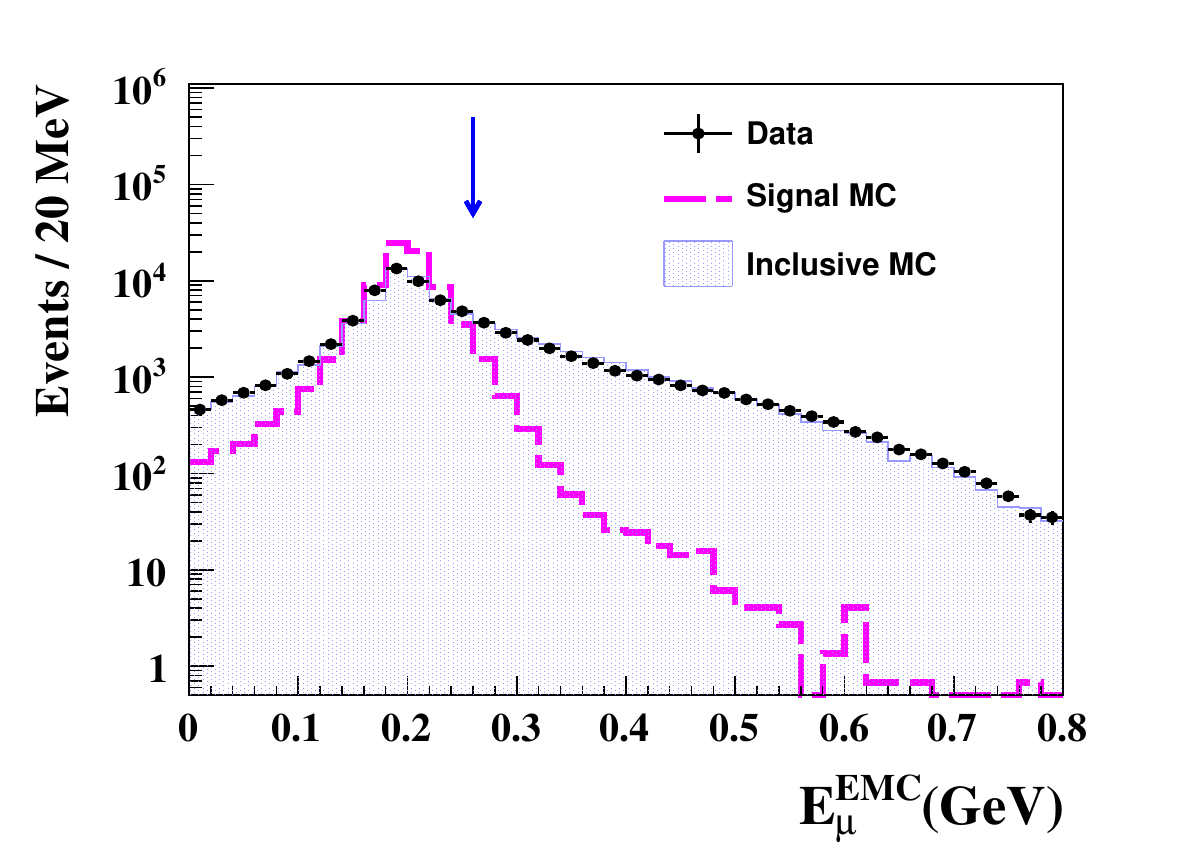}}
        \subfigure[]
        {\includegraphics[width=0.49\textwidth]{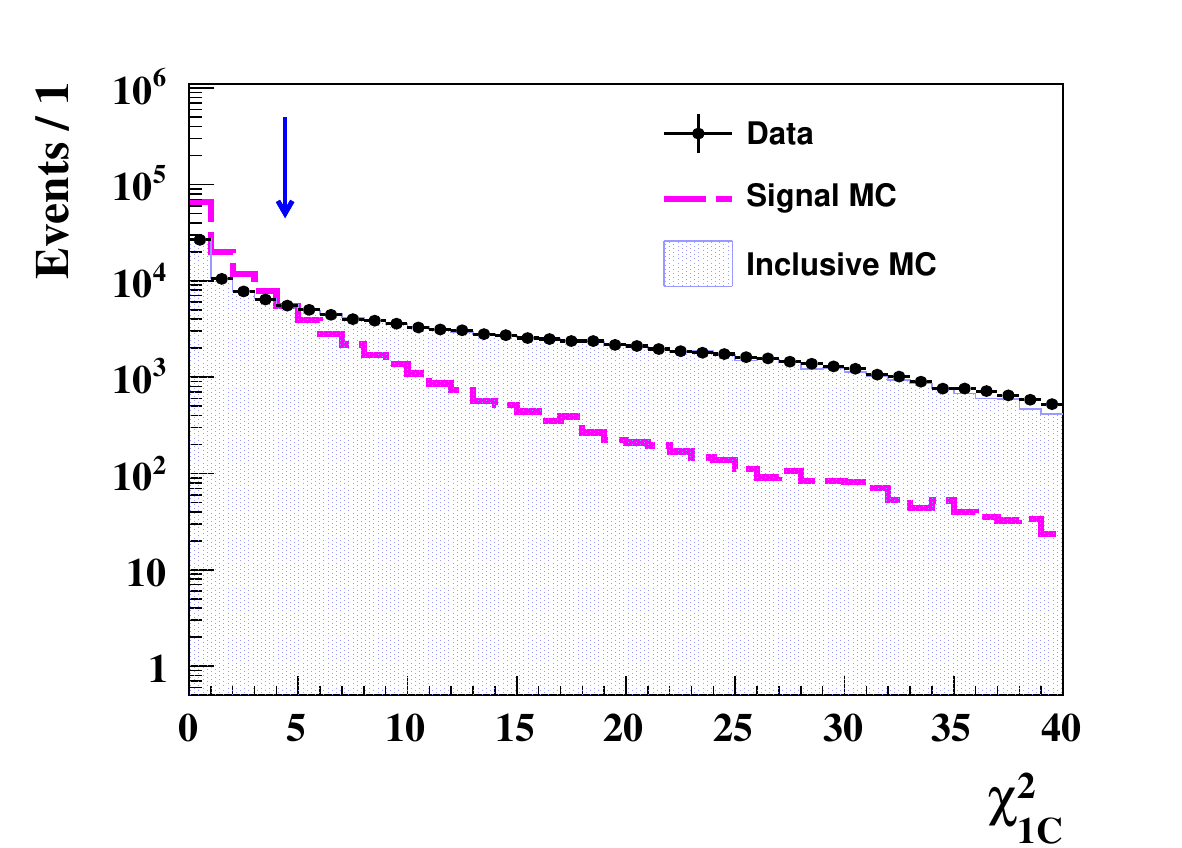}}\\
        
	\caption{Distributions of $E_{\mu}^{\rm{EMC}}$ (a) and $\chi_{1C}^{2}$ (b) from data, signal MC sample and inclusive MC sample. The black dots with error bars are data, the blue shaded histogram is the inclusive MC sample and the magenta dashed line is the signal MC sample. In plot (a), the blue arrow indicates $E_{\mu}^{\rm{EMC}}<0.26\gev$ for muon candidates. In plot (b), the blue arrow indicates $\chi_{1C}^{2}<4.4$ for $D^-$ meson candidates.} 
	\label{fig:cut}
\end{figure*}

\hspace{1.5em} 
The analysis is performed with the BESIII offline software system~\cite{bes3:boss705}.
In the signal process $\jpsi\to D^{-}\mu^{+}\nu_{\mu}$, $D^{-}\to K^{+}\pi^{-}\pi^{-}$, 
all final-state particles except the $\nu_{\mu}$ are detected.
Charged tracks detected in the MDC are required to be within a polar angle ($\theta$) range of $|\rm{cos\theta}|<0.93$, where $\theta$ is defined with respect to the $z$ axis, which is the symmetry axis of the MDC. 
For all charged tracks, the distance of closest approach to the interaction point (IP) 
must be less than 10\,cm along the $z$ axis and less than 1\,cm in the transverse plane. 
Only events with exactly four selected charged tracks and zero net charge are retained for further analysis. 
Particle identification~(PID) for charged tracks combines measurements of the energy deposited in the MDC~(d$E$/d$x$) and the flight time in the TOF to form likelihoods $\mathcal{L}(h)~(h=K,\pi)$ for each hadron $h$ hypothesis.
Charged kaons are identified by $\mathcal{L}(K)>0$ and $\mathcal{L}(K)>\mathcal{L}(\pi)$, while charged pions are identified by $\mathcal{L}(\pi)>0$ and $\mathcal{L}(\pi)>\mathcal{L}(K)$.
Muon PID uses information measured in the MDC, TOF and EMC. 
Most muons in $\jpsi\to \dmunv+c.c.$ are low enough in momentum that using the MUC for muon identification is ineffective because of low detection efficiency.
The combined likelihoods ($\mathcal{L}'$) under the muon, electron (positron), and kaon hypotheses are obtained. Muon candidates are required to satisfy $\mathcal{L}'(\mu)>0.001$, $\mathcal{L}'(\mu)>\mathcal{L}'(e)$ and $\mathcal{L}'(\mu)>\mathcal{L}'(K)$.
To further reduce the background, the deposited energy of the muon candidate in the EMC, $E_{\mu}^{\rm{EMC}}$, is required to be less than $0.26 \gev$, as shown in Fig.~\ref{fig:cut} (a). 

\vspace{-0.0cm}
\vspace{-0.0cm}

Photon candidates are identified using showers in the EMC.  The deposited energy of each shower must exceed 25~MeV in the barrel region ($|\cos \theta|< 0.80$) and 50~MeV in the end cap region ($0.86 <|\cos \theta|< 0.92$).
To suppress electronic noise and showers unrelated to the event, the difference between the EMC shower time and the event start time is required to be within [0, 700]\,ns.
In addition, 
to exclude showers that originate from charged tracks, the angle subtended by the EMC shower and the position of the closest charged track at the EMC must be greater than 10 degrees as measured from the IP. 
Because the signal channel has no photon in the final state, we require the total energy of good photons ($E_{\gamma}^{\rm{tot}}$) to be less than $0.1\gev$ to suppress backgrounds with a $\pi^{0}$ or extra photon(s). 

The $D^{-}$ meson is reconstructed through the process $D^{-}\to K^+\pi^-\pi^-$, and its invariant mass $M_{K\pi\pi}$ is required to be in the range of $[1.85, 1.89]\gevcc$, corresponding to $\pm3$ times the mass resolution around the known $D^-$ mass~\cite{pdg:2022}.
In addition, a kinematic fit constraining $M_{K\pi\pi}$ to the known $D^-$ mass is performed and the obtained $\chi^2_{\rm{1C}}$ value is required to be less than 4.4 for $D^-$ meson candidates, where $\chi^2_{\rm{1C}}$ is obtained from lagrange multiplier method and a smaller $\chi^2_{\rm{1C}}$ value indicates a better fit~\cite{chi2}, as shown in Fig.~\ref{fig:cut} (b). The momentum of the neutrino $\nu_\mu$, which is not detected, is inferred from the missing momentum $|\vec{P}_{\rm{miss}}|$ defined as
 \begin{eqnarray}
 |\vec{P}_{\rm{miss}}|=|\vec{0}-\vec{P}_{D^-}-\vec{P}_{\mu^+}|, 
 \end{eqnarray}
where $\vec{P}_{D^-}$ ($\vec{P}_{\mu^+}$) is the momentum of the $D^-$ ($\mu^+$) in the rest frame of the initial $e^+e^-$ collision, and $|\vec{P}_{\rm{miss}}|$ is required to be greater than $0.05\gevc$ to reduce backgrounds without missing particles. Furthermore, to suppress backgrounds from non-three-body decays, we require $0.98 \gevc<|\vec{P}_{\mu^+}|+|\vec{P}_{\rm{miss}}|<1.23 \gevc$, a range including $>95\%$ of the signal events (with limits from colinear decay configurations).  

After the above selection criteria, several hadronic backgrounds still exist in the inclusive MC samples.  
The main backgrounds are composed of the following four sources~\cite{ref:topoana}. 
\begin{itemize}
\item $\jpsi\to K^+K_S\pi^-$ and $\jpsi\to K^+K_S\pi^-\pi^0$ with $K_S\to\pi^+\pi^-$, where one of the pions from the $K_S$ is misidentified as muon. 
The pair mass of oppositely-charged pions $M_{\pi\pi}$ closest to the known $K_S$ mass~\cite{pdg:2022} is required to be greater than $0.52\gevcc$ to veto these backgrounds. 
A lower cut on the mass is omitted to also veto events with a $K_S$ decay product undergoing $\pi^\pm \to \mu^\pm \nu_\mu$ decay.  
\item $\jpsi\to K^+K^-\pi^+\pi^-$ background, in which one of kaons decays through $K^-\to\mu^-\bar{\nu}_{\mu}$ or $K^-\to\pi^0\mu^-\bar{\nu}_{\mu}$, where the muon is misidentified as pion and one of pions is misidentified as muon (double misidentification is needed to obtain like-sign pion candidates). To reject these backgrounds, we calculate the momentum of the decay-kaon $K^-$ by $\vec{P}_{K^-}=\vec{P}_{\jpsi}-\vec{P}_{K^+}-\vec{P}_{\pi^+}-\vec{P}_{\pi^-}$ and reconstruct the invariant mass $M_{2K2\pi}$ of $\jpsi\to K^+K^-\pi^+\pi^-$. Events with $M_{2K2\pi}$ in $[3.07,3.13]\gevcc$ are vetoed. 
\item $\jpsi\to \pi^+\pi^-\pi^+\pi^-\pi^0$ background, where one of the pions is misidentified as muon, another pion is misidentified as kaon and the $\pi^0$ contributes to the missing momentum. The momentum of the $\pi^0$ can be calculated by $\vec{P}_{\pi^0}=\vec{P}_{\jpsi}-\vec{P}_{\pi^+}-\vec{P}_{\pi^+}-\vec{P}_{\pi^-}-\vec{P}_{\pi^-}$ and the invariant mass $M_{4\pi\pi^0}$ of $\jpsi\to \pi^+\pi^-\pi^+\pi^-\pi^0$ is required to be less than $3.05\gevcc$ to reduce this background. 
\item $\jpsi\to K^+\pi^-\pi^+\pi^-K_L$ background, where $K_L$ is a missing particle. Similar to $\jpsi\to \pi^+\pi^-\pi^+\pi^-\pi^0$, we deduce the momentum of $K_L$ from $\vec{P}_{K_L}=\vec{P}_{\jpsi}-\vec{P}_{K^+}-\vec{P}_{\pi^+}-\vec{P}_{\pi^-}-\vec{P}_{\pi^-}$. The invariant mass $M_{K3\pi K_L}$ of $\jpsi\to K^+\pi^-\pi^+\pi^-K_L$ is required to be greater than $3.24\gevcc$ to suppress this background.  Note that $M_{K3\pi K_L}$ calculated for the signal mode is increased due to the $K_L$ mass, and raising the requirement well above the known $\jpsi$ mass~\cite{pdg:2022} removes some other miscellaneous backgrounds without decreasing signal efficiency.  
\end{itemize}
The number of main backgrounds before and after vetoes are shown in Table~\ref{tab:bkg}, which is estimated from MC samples. Most of the expected backgrounds have been vetoed, and the estimated remaining backgrounds are not subtracted from the result. These vetoes could further improve the sensitivity of detecting the rare process.

\begin{table*}[!htbp]
\caption{The number of main backgrounds from inclusive MC before (the second column) and after (the third column) vetoes. The fourth column shows the corresponding veto ratio. }
\setlength{\abovecaptionskip}{1.2cm}
\setlength{\belowcaptionskip}{0.2cm}
\begin{center}
\footnotesize
\vspace{-0.0cm}
\begin{tabular}{lccc}
\hline \hline
                    Backgrounds & Before vetoes  & After vetoes & Veto ratio\\
                    \hline
                    $\jpsi\to K^+\pi^-\pi^+\pi^-(\pi^0)$ & 12456 & 91 & 99\%\\
                    $\jpsi\to K^+K^-\pi^+\pi^-$ & 10066 & 459 & 95\%\\
                    $\jpsi\to \pi^+\pi^-\pi^+\pi^-\pi^0$ & 13694 & 332 & 98\%\\
                    $\jpsi\to K^+\pi^-\pi^+\pi^-K_L$ & 3852 & 150 & 96\%\\
    
\hline \hline
\end{tabular}
\label{tab:bkg}
\end{center}
\end{table*}
\vspace{-0.0cm}

\begin{figure}[htbp] \centering
	\setlength{\abovecaptionskip}{-1pt}
	\setlength{\belowcaptionskip}{10pt}
	\includegraphics[width=10.0cm]{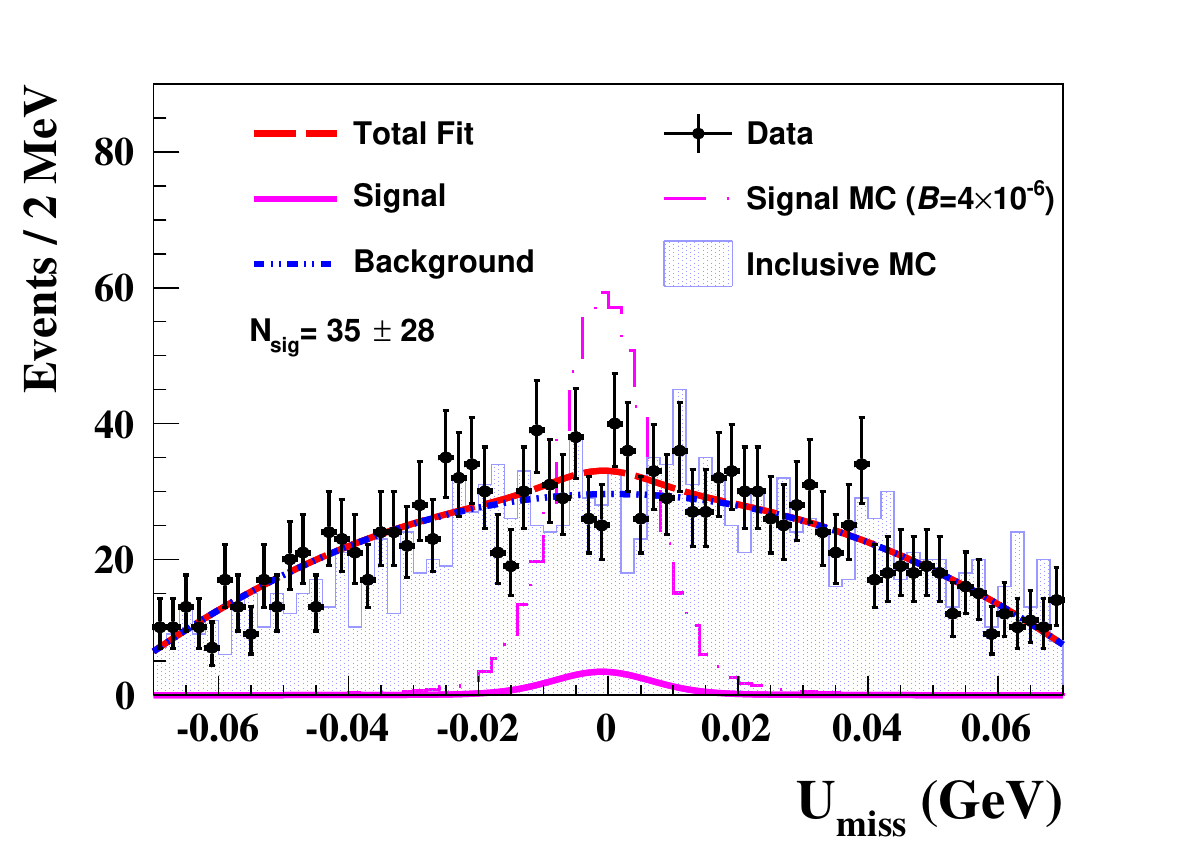}
	\caption{The distribution of $U_{\rm{miss}}$ of the accepted candidates in data, signal MC sample and inclusive MC sample. The black dots with error bars are data, the blue shaded histogram is the inclusive MC sample, and the magenta dotted-dashed line is the signal MC sample. The red dashed line is the total fit result, the magenta solid line is the signal from fit and the blue dotted-dashed line is the background from fit. Here, the signal MC sample is drawn with the assumption of $\mathcal{B}(\jpsi\to D^{-}\mu^{+}\nu_{\mu})=4\times10^{-6}$.}
	\label{fig:fit}
\end{figure}

We extract the signal yield of $\jpsi\to\dmunv$ by examining the kinematic variable
 \begin{eqnarray}
 U_{\rm{miss}}=E_{\rm{miss}}-|\vec{P}_{\rm{miss}}|c, 
 \end{eqnarray}
where $E_{\rm{miss}}$ is the missing energy calculated by $E_{\rm{miss}}=E_{\jpsi}-E_{D^-}-E_{\mu^+}$ and $E_{D^-}$ ($E_{\mu^+}$) is the energy of $D^-$ ($\mu^+$) in the rest frame of the initial $e^+e^-$ collision. 
In such a distribution, the signal channel would appear as a Gaussian with its centroid at 0, as shown by the magenta dotted-dashed line in Fig.~\ref{fig:fit}.

\vspace{-0.0cm}
\vspace{-0.0cm}

\section{RESULT}
\label{sec:result}
\hspace{1.5em}
To extract the signal yield, an unbinned extended maximum likelihood fit in $U_{\rm{miss}}$ is performed for data. The fit function is defined as
\begin{eqnarray}
\label{eq:fit umiss}
\mathcal{F}_{\rm{total}} = N_{\rm{sig}}{\times}\mathcal{PDF}_{\rm{sig}}\bigotimes G(\mu,\sigma)+N_{\rm{bkg}}{\times}Poly(c_{0},c_{1}),
\end{eqnarray}
with each term defined as follows:
\begin{itemize}
\item $N_{\rm{sig}}$: The number of signal events.
\item $N_{\rm{bkg}}$: The number of background events.
\item $\mathcal{PDF}_{\rm{sig}}\bigotimes G(\mu,\sigma)$: The probability density function derived from the shape of signal MC simulation of $U_{\rm{miss}}$ spectrum convolved with a Gaussian function $G(\mu,\sigma)$, where $\mu$ and $\sigma$ are obtained from the control sample.
\item $G(\mu,\sigma)$: The Gaussian function with $\mu=(0.35\pm0.12)\mev$ and $\sigma=(3.32\pm0.27)\mev$ to describe the resolution difference and mean shift of the signal between data and MC simulation. We perform a fit to the control sample $D^0\to K^{-}\mu^{+}\nu_{\mu}$ from $\psi(3773)\to D^0\bar{D}^0$ data with $G(\mu,\sigma)$ floating to obtain the values of $\mu$ and $\sigma$.
\item $Poly(c_{0},c_{1})$: A second-order polynomial shape to describe the backgrounds.
\end{itemize}
We try different signal regions and background shapes to calculate the UL, and the option with the most conservative UL is chosen: [-0.07, 0.07] GeV for the signal region and a second-order polynomial for the background shape.

As shown in Fig.~\ref{fig:fit}, the fit returns $N_{\rm{sig}}=35\pm28$, showing no significant excess of signal above the background.
The BF of the signal decay is calculated by
\begin{eqnarray}
\mathcal{B}(\jpsi\to D^{-}\mu^{+}\nu_{\mu})=\frac{N_{\rm{sig}}}{N_{\jpsi} \,\cdot\, \hat{\epsilon} \,\cdot\, \mathcal{B}_{\rm{sub}}},
\end{eqnarray}
where $N_{\jpsi}=(10087\pm44)\times10^{6}$ is the total number of $\jpsi$ events~\cite{bes3:njpsi2022}, $\hat{\epsilon}=(14.29\pm0.05)\%$ is the signal detection efficiency, where the uncertainty is statistical only, and $\mathcal{B}_{\rm{sub}}=(9.38\pm0.16)\%$ is the BF of the intermediate decay $D^{-}\to K^{+}\pi^{-}\pi^{-}$~\cite{pdg:2022}. 

\begin{figure}[tp] \centering
	\setlength{\abovecaptionskip}{-1pt}
	\setlength{\belowcaptionskip}{10pt}
	\includegraphics[width=10.0cm]{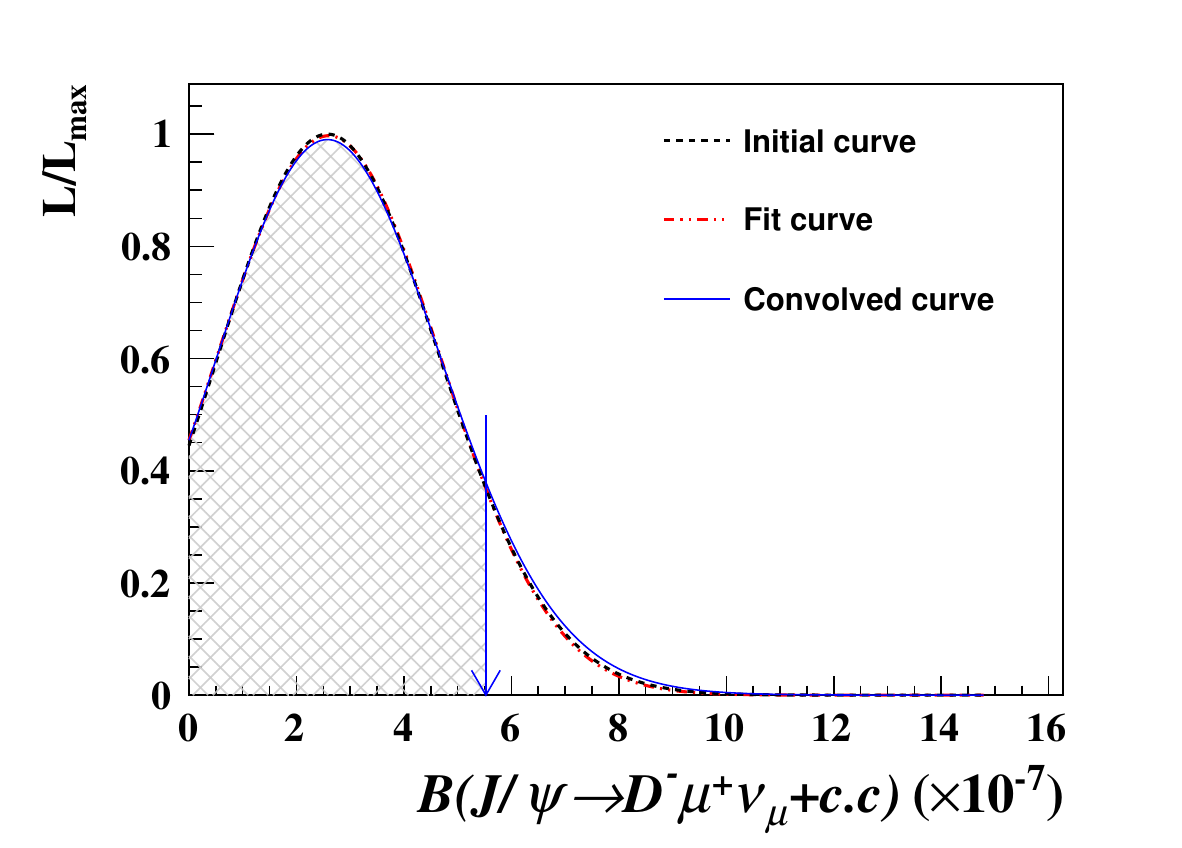}
	\caption{The distribution of likelihood scan values. The black dotted-dashed line is the initial curve, the red dotted-dashed line is the fit of the initial curve with a Gaussian function (fit curve) and the blue solid line is the result after convolution with a Gaussian for systematic uncertainties (convolved curve). The blue arrow indicates the UL on the BF at the 90\% C.L.}
	\label{fig:UL}
\end{figure}

Since no significant excess of signal above the background is observed, an upper limit (UL) on the BF is set with a Bayesian approach. A series of fits are performed, where the number of signal events $N_{\rm{sig}}$ are fixed to values from 0 to 200 in steps of 0.1.
For each fixed signal number, a calculated BF value and a likelihood value are obtained by fit (initial curve; see Fig.~\ref{fig:UL}).
Then we fit the likelihood values as a Gaussian function of the BFs (fit curve):
\begin{eqnarray}
\mathcal{L(B)}_{\rm{fit}}\varpropto \rm{exp}\left[- \, \frac{(\mathcal{B}-\hat{\mathcal{B}})^{2}}{2\sigma_{\mathcal{B}}^{2}}\right],
\label{eq:lh}
\end{eqnarray}
where $\hat{\mathcal{B}}$ is the mean value and $\sigma_{\mathcal{B}}$ is the sigma value of the Gaussian function from fit.
To include the systematic uncertainties described in the following section, we follow the method in Refs.~\cite{Liu:2015uha,UL:smear} of convolving with a Gaussian function:
\begin{eqnarray}
\mathcal{L(B)}_{\rm{smear}}\varpropto\int_{0}^{1}\rm{exp}\left[- \, \frac{(\epsilon\mathcal{B}/\hat\epsilon-\hat{\mathcal{B}})^{2}}{2\sigma_{\mathcal{B}}^{2}}\right]\times \frac{1}{\sqrt{2\pi}\sigma_{\epsilon}} \, \rm{exp}\left[- \, \frac{(\epsilon-\hat{\epsilon})^{2}}{2\sigma_{\epsilon}^{2}}\right]~d\epsilon,
\label{eq:smear}
\end{eqnarray}
where $\hat\epsilon$ is the nominal efficiency and $\sigma_{\epsilon}=\Delta_{\rm{syst}}\cdot\hat\epsilon$ ($\Delta_{\rm{syst}}$ is the relative systematic uncertainty which is further discussed in Section~\ref{sec:systematic}) is the systematic uncertainty on the efficiency.
The distributions of the likelihood curves are shown in Fig.~\ref{fig:UL}.
The UL on the BF at 90\% confidence level is obtained by integrating the convolved likelihood curve from zero to $90\%$ in the physical region ($\mathcal{B}\geq0$), resulting in $\mathcal{B}(\jpsi\to D^{-}\mu^{+}\nu_{\mu})<5.6\times10^{-7}$ at $90\%$ C.L.

\section{SYSTEMATIC UNCERTAINTY}
\label{sec:systematic}
\hspace{1.5em}
The systematic uncertainties in measuring the BF of $\jpsi\to D^{-}\mu^{+}\nu_{\mu}$ come from the signal MC model, the value of $G(\mu,\sigma)$, tracking and PID efficiencies, 
the $D^-$ decay BF, the total number of $\jpsi$ events, 
and event selection requirements.  For the latter case, data-MC differences measured with control samples are used to evaluate uncertainties.  
The systematic uncertainties are described next and are summarized in Table~\ref{tab:syst_err}.  

\begin{table*}[tpb]
\setlength{\abovecaptionskip}{0.0cm}
\setlength{\belowcaptionskip}{-1.6cm}
\caption{Summary of the systematic uncertainties for the measurement of the BF. The total value is calculated by summing up all sources in quadrature.}
  \begin{center}
  \footnotesize
  \newcommand{\tabincell}[2]{\begin{tabular}{@{}#1@{}}#2\end{tabular}}
  \begin{threeparttable}
  \begin{tabular}{l|c c c c c c c}
      \hline\hline
                         Sources & Relative uncertainties ($\times10^{-2}$)\\
			\hline
	    Signal model & 5.8\\
		Tracking & 4.0\\
		PID & 3.5\\
		Limited MC sample  & 0.4\\
		$\mathcal{B}(D^{-}\to K^{+}\pi^{-}\pi^{-})$ & 1.7\\
		Total number of $J/\psi$ & 0.5\\
		$M_{K\pi\pi}$ requirement & 0.6\\
		$\chi_{1C}^2$ requirement & 3.5\\
		$|\vec{P}_{\mu}|+|\vec{P}_{\rm{miss}}|$ requirement & 1.7\\
		$E_{\gamma}^{\rm{tot}}$ requirement & 1.7\\
		$M_{\pi\pi}$ requirement & 1.0\\
		$M_{2K2\pi}$ requirement & 1.1\\
		$M_{K3{\pi}K_{L}}$ requirement & 2.1\\
    			\hline
                         Total & 9.5\\
      \hline\hline
  \end{tabular}
  \label{tab:syst_err}
  \end{threeparttable}
  \end{center}
\end{table*}

\begin{itemize}
    \item \emph{Signal model.} To estimate the systematic uncertainty due to the signal model, we use a phase space (PHSP) model. The 5.8\% difference between the efficiencies of the nominal and PHSP models is assigned as the systematic uncertainty. 

    \item \emph{The value of $G(\mu,\sigma)$.} The uncertainty of $\mu$ and $\sigma$ in $G(\mu,\sigma)$ are estimated from the control sample $D^0\to K^{-}\mu^{+}\nu_{\mu}$ (Section \ref{sec:result}). To consider the uncertainty of $G(\mu,\sigma)$, the values of $\mu$ and $\sigma$ are assumed to follow a Gaussian distribution and constrained by multiplying the likelihood function with two additional Gaussian functions in the fit. The contribution of this item is found to be less than 0.1\%, which is negligible to the total systematic uncertainty.

    \item \emph{Tracking and PID efficiencies.} The uncertainty due to tracking and PID efficiencies for kaons and pions are estimated with the control samples of $D^{+}\to K^{-}\pi^{+}\pi^{+}$ and $D^{-}\to K^{+}\pi^{-}\pi^{-}$ from $\psi(3773)\to D^+D^-$, where one $\pi^{-}$ or $K^{+}$ meson is not reconstructed~\cite{bes3:kpi2017}.  The uncertainties of tracking (PID) are estimated to be 1.0\% (1.0\%) per track. 
    In addition, the uncertainties of tracking and PID efficiencies from the muon, estimated by studying a control sample of $\ee\to\gamma\mu^+\mu^-$, are 1.0\% for tracking and 0.5\% for PID.

     \item \emph{Limited signal MC sample.} The signal efficiency is estimated from signal MC events, and the MC statistics is considered to be one source of systematic uncertainty, which is 0.4\%.

     \item \emph{BF of $D^{-}\to K^{+}\pi^{-}\pi^{-}$ decay.} The uncertainty on $\mathcal{B}(D^{-}\to K^{+}\pi^{-}\pi^{-})$ is $1.7\%$~\cite{pdg:2022}. 

     \item \emph{Number of $\jpsi$ events.} We take the relative uncertainty of $0.5\%$ determined using $\jpsi$ inclusive hadronic decays for $N_{\jpsi}$ as the systematic uncertainty~\cite{bes3:njpsi2022}.

    \item \emph{$M_{K\pi\pi}$ and $\chi_{1C}^2$ requirements.} The control sample of $\psi(3770)\to D^+D^-,D^-\to K^+\pi^-\pi^-$ is used to estimate the systematic uncertainties of the $M_{K\pi\pi}$ and $\chi_{1C}^2$ requirements as 0.6\% and 3.5\%, respectively.

    \item \emph{$|\vec{P}_{\mu}|+|\vec{P}_{\rm{miss}}|$ requirements.} The control sample of $\psi(3770)\to D^+D^-\to D^+\pi^{-}K_{S}$ is used to estimate the systematic uncertainties of $|\vec{P}_{\mu}|+|\vec{P}_{\rm{miss}}|$ requirements, where the $K_S$ is considered as the missing particle. The uncertainty is assigned as 1.7\%. 

    \item \emph{$E_{\gamma}^{\rm{tot}}$, $|\vec{P}_{\rm{miss}}|$ requirements.} The control sample $D^0\to K^{-}\mu^{+}\nu_{\mu}$ from $\psi(3773)\to D^0\bar{D}^0$ data is studied. The uncertainty of $E_{\gamma}^{\rm{tot}}$ requirement is 1.7\%, and the uncertainty of $|\vec{P}_{\rm{miss}}|$ requirement is negligible.

    \item \emph{$M_{\pi\pi}$ veto.} The control sample of $\jpsi\to K^+K^-\pi^+\pi^-$, which cannot include the decay $K_S\to\pi^+\pi^-$ due to strangeness conservation, is used to estimate the systematic uncertainty of $M_{\pi\pi}$ veto, which is 1.0\%.

    \item \emph{$M_{2K2\pi}$, $M_{4\pi\pi^0}$ and $M_{K3\pi K_L}$ vetoes.} $\jpsi\to K^+\pi^-\pi^+\pi^-\pi^0$ is selected as the control sample for study, where the $\pi^0$ is considered as the missing particle. The systematic uncertainties of the $M_{2K2\pi}$, and $M_{K3\pi K_L}$ vetoes are estimated to be 1.1\% and 2.1\% respectively, and the systematic uncertainty of the $M_{4\pi\pi^0}$ veto is negligible.

\end{itemize}
The sum of all contributions in quadrature, $9.5\%$, is taken as the total systematic uncertainty.  


\section{SUMMARY}
\label{sec:summary}
\hspace{1.5em}
Based on $(10087\pm44)\times10^{6}$ $\jpsi$ events collected with the BESIII detector,
the BF of rare semi-leptonic decay $\jpsi\to D^{-}\mu^{+}\nu_{\mu}$ is searched for. 
No significant signal is observed.
The UL on the BF is set to be $\mathcal{B}(\jpsi\to D^{-}\mu^{+}\nu_{\mu})<5.6\times10^{-7}$ at 90\% confidence level, where systematic uncertainties are taken into account. 
This is the first search of a charmonium weak decay with a muon in the final state. The result is compatible with the SM based predictions~\cite{wang:2008b, shen:2008, dhir:2013, ivanov:2015, tian:2017}, and the measurement is not sensitive to the predicted level by 4 orders of magnitude.

\acknowledgments
\hspace{1.5em}
The BESIII Collaboration thanks the staff of BEPCII and the IHEP computing center for their strong support. This work is supported in part by National Key R\&D Program of China under Contracts Nos. 2020YFA0406400, 2020YFA0406300; National Natural Science Foundation of China (NSFC) under Contracts Nos. 11635010, 11675275, 11735014, 11835012, 11935015, 11935016, 11935018, 11961141012, 11975021, 12022510, 12025502, 12035009, 12035013, 12061131003, 12175321, 12192260, 12192261, 12192262, 12192263, 12192264, 12192265, 12221005, 12225509, 12235017; the Chinese Academy of Sciences (CAS) Large-Scale Scientific Facility Program; the CAS Center for Excellence in Particle Physics (CCEPP); Joint Large-Scale Scientific Facility Funds of the NSFC and CAS under Contract No. U1832207, U1932101; CAS Key Research Program of Frontier Sciences under Contracts Nos. QYZDJ-SSW-SLH003, QYZDJ-SSW-SLH040; 100 Talents Program of CAS; State Key Laboratory of Nuclear Physics and Technology, PKU under Grant No. NPT2020KFY04; The Institute of Nuclear and Particle Physics (INPAC) and Shanghai Key Laboratory for Particle Physics and Cosmology; ERC under Contract No. 758462; European Union's Horizon 2020 research and innovation programme under Marie Sklodowska-Curie grant agreement under Contract No. 894790; German Research Foundation DFG under Contracts Nos. 443159800, 455635585, Collaborative Research Center CRC 1044, FOR5327, GRK 2149; Istituto Nazionale di Fisica Nucleare, Italy; Ministry of Development of Turkey under Contract No. DPT2006K-120470; National Research Foundation of Korea under Contract No. NRF-2022R1A2C1092335; National Science and Technology fund of Mongolia; National Science Research and Innovation Fund (NSRF) via the Program Management Unit for Human Resources \& Institutional Development, Research and Innovation of Thailand under Contract No. B16F640076; Polish National Science Centre under Contract No. 2019/35/O/ST2/02907; The Swedish Research Council; U. S. Department of Energy under Contract No. DE-FG02-05ER41374.

\newpage
\input{authorlist_2023-03-11.tex}
\end{document}

%% file: authorlist_2023-03-11.tex
\begin{small}
\begin{center}
M.~Ablikim$^{1}$, M.~N.~Achasov$^{5,b}$, P.~Adlarson$^{75}$, X.~C.~Ai$^{81}$, R.~Aliberti$^{36}$, A.~Amoroso$^{74A,74C}$, M.~R.~An$^{40}$, Q.~An$^{71,58}$, Y.~Bai$^{57}$, O.~Bakina$^{37}$
% ...
\\
\vspace{0.2cm}
(BESIII Collaboration)\\
\vspace{0.2cm} {\it
$^{1}$ Institute of High Energy Physics, Beijing 100049, People's Republic of China\\
% ...
$^{5}$ Budker Institute of Nuclear Physics SB RAS (BINP), Novosibirsk 630090, Russia\\
% ...
$^{36}$ Johannes Gutenberg University of Mainz, Johann-Joachim-Becher-Weg 45, D-55099 Mainz, Germany\\
$^{37}$ Joint Institute for Nuclear Research, 141980 Dubna, Moscow region, Russia\\
% ...
$^{40}$ Liaoning Normal University, Dalian 116029, People's Republic of China\\
% ...
$^{57}$ Southeast University, Nanjing 211100, People's Republic of China\\
$^{58}$ State Key Laboratory of Particle Detection and Electronics, Beijing 100049, Hefei 230026, People's Republic of China\\
% ...
$^{71}$ University of Science and Technology of China, Hefei 230026, People's Republic of China\\
% ...
$^{74}$ University of Turin and INFN, (A)University of Turin, I-10125, Turin, Italy; (B)University of Eastern Piedmont, I-15121, Alessandria, Italy; (C)INFN, I-10125, Turin, Italy\\
$^{75}$ Uppsala University, Box 516, SE-75120 Uppsala, Sweden\\
% ...
$^{81}$ Zhengzhou University, Zhengzhou 450001, People's Republic of China\\

\vspace{0.2cm}
% ...
$^{b}$ Also at the Novosibirsk State University, Novosibirsk, 630090, Russia\\
% ...

}
\end{center}

\vspace{0.4cm}
\end{small}